\documentstyle[aps,psfig,subfigure]{revtex}
\tightenlines

\begin{document}
\draft

\title{Mean-field theory of learning: from dynamics to statics}
\author{K. Y.~Michael Wong, S.~Li and Peixun~Luo}
\address{
Department of Physics, Hong Kong University of Science and Technology,
Clear Water Bay, Kowloon, Hong Kong
}
\date{June 14, 2000}
\maketitle


\begin{abstract}
Using the cavity method and diagrammatic methods, 
we model the dynamics of batch learning
of restricted sets of examples.
Simulations of the Green's function
and the cavity activation distributions
support the theory well.
The learning dynamics approaches a steady state
in agreement with the static version of the cavity method.
The picture of the rough energy landscape is reviewed.
\end{abstract}

\section{Introduction} 

The mean-field theory was first developed as an {\it approximation}
to many physical systems in magnetic or disordered materials
\cite{statphys}.
However, it is interesting that they become {\it exact}
in many systems in information processing.
The major reason of its success
is that when compared with physical systems,
these artificial systems have extensive interactions
among their components.
Hence when one component is considered,
the influence of the rest of the system can be regarded
as a background satisfying some averaged properties.

Learning in large neural networks is a mean-field process
since the examples and weights strongly interact with each other
during the learning process.
Learning is often achieved by defining an energy function
which involves a training set of examples.
The energy function is then minimized
by a gradient descent process with respect to the weights
until a steady state is reached.
Each of the many weights is thus dependent
on each of the many examples and vice versa.
This makes it an ideal area
for applying mean-field theories.

There have been attempts using mean-field theories 
to describe the dynamics of learning.
In batch learning, the same restricted set of examples
is provided for {\it each} learning step.
Using the dynamical mean field theory,
early work has been done
on the steady-state behavior and asymptotic time scales 
in perceptrons with binary weights,
rather than the continuous weights of more common interest \cite{horner}.
Much benchmarking of batch learning has been done for linear learning rules
such as Hebbian learning \cite{coolen,rae}
or Adaline learning \cite{hertz}.
The work on Adaline learning was further extended
to the study of linear perceptrons learning nonlinear rules \cite{bos,bos2}.
However, not much work has been done
on the learning of nonlinear rules with continuous weights.
In this respect, it is interesting to note the recent attempts
using the {\it dynamical replica theory} \cite{coolen,rae}.
It approximates the temporal correlations during learning
by instantaneous effective macroscopic variables.
Further approximations facilitate results for nonlinear learning.
However, the rigor of these approximations
remain to be confirmed in the general case.

Batch learning is different from idealized models
of {\it on-line} learning of infinite training sets,
which has gained much progress
\cite{biehl,saadsolla1,saadsolla2,saadrattray,online}.
In this model, an independent example is generated for each learning step.
Since statistical correlations among the examples can be ignored, 
the many-body interactions among the examples,
and hence among the weights, are absent.
Hence they do not address the many-body aspects of the dynamics,
which will be discussed here.
Nevertheless, this simplification enables the dynamics
to be simply described by instantaneous dynamical variables,
resulting in a significant reduction
in the complexity of analysis,
thereby leading to great advances in our understanding of on-line learning.
In multilayer perceptrons, for instance,
the persistence of a permutation symmetric stage
which retards the learning process was well studied.
Subsequent proposals to speed up learning were made,
illustrating the usefulness of the on-line approach
\cite{saadrattray,natural}.

Here we review models of batch learning \cite{many,nips99}
where, however, such simplifications are not available.
Since the same restricted set of examples is recycled
during the learning process,
there now exist temporal correlations of the parameters
in the learning history.
Nevertheless, we manage to consider
the learning model as a many-body system. 
Each example makes a small contribution to the learning process,
which can be described by linear response terms 
in a sea of background examples. 
Two ingredients are important to our theory:

(a) {\it The cavity method} --
Originally developed as the Thouless-Anderson-Palmer approach 
to magnetic systems and spin glasses \cite{mpv},
the method was adopted to learning in perceptrons \cite{mezard},
and subsequently extended to the teacher-student perceptron \cite{bouten},
the AND machine \cite{grini},
the multiclass perceptron \cite{gerl},
the committee tree \cite{epl,nips96},
Bayesian learning \cite{opperwinther}
and pruned perceptrons \cite{tanc}.
These studies only considered the equilibrium properties of learning,
whereas here we are generalizing the method
to study the dynamics \cite{mpv}.
It uses a self-consistency argument
to compare the evolution of the activation of an example
when it is absent or present in the training set.
When absent, the activation of the example
is called the {\it cavity activation},
in contrast to its generic counterpart
when it is included in the training set.

The cavity method yields macroscopic properties
identical to the more conventional replica method \cite{mpv}.
However, since the replica method was originally devised
as a technique to facilitate systemwide averages,
it provides much less information
on the microscopic conditions of the individual dynamical variables.

(b) {\it The diagrammatic approach} --
To describe the difference between the cavity activation
and its generic counterpart of an example,
we apply linear response theory
and use Green's function to describe
how the influence of the added example 
propagates through the learning history.
The Green's function is represented by a series of diagrams,
whose averages over examples are performed by a set of pairing rules
similar to those introduced for Adaline learning \cite{hertz},
as well as in the dynamics of layered networks \cite{layer}.
Here we take a further step and use the diagrams to describe
the changes from cavity to generic activations,
as was done in \cite{exact},
rather than the evolution of specific dynamical variables
in the case of linear rules \cite{hertz}. 
Hence our dynamical equations are widely applicable
to any gradient-descent learning rule
which minimizes an {\it arbitrary} cost function 
in terms of the activation. 
It fully takes into account the temporal correlations during learning, 
and is exact for large networks. 

The study of learning dynamics should also provide further insights
on the steady-state properties of learning.
In this respect we will review the cavity approach
to the steady-state behavior of learning,
and the microscopic variables satisfy a set of TAP equations.
The approach is particularly transparent
when the energy landscape is smooth,
i.e., no local minima interfere with the approach to the steady state.
However, the picture is valid
only when a stability condition
(equivalent to the Almeida-Thouless condition in the replica method)
is satisfied.
Beyond this regime, local minima begin to appear
and the energy landscape is roughened.
In this case, a similar set of TAP equations remains valid.
The physical picture has been presented in \cite{nips96};
a more complete analysis is presented here.

The paper is organized as follows.
In Section 2 we formulate the dynamics of batch learning.
In Section 3 we introduce the cavity method
and the dynamical equations for the macroscopic variables.
In Section 4 we present simulation results which support the cavity theory.
In Sections 5 and 6 we consider the steady-state behaviour of learning
and generalize the TAP equations
respectively to the pictures of smooth and rough energy landscapes,
followed by a conclusion in Section 7.
The appendices explain the diagrammatic approach
in describing the Green's function,
the fluctuation response relation,
and the equations for macroscopic parameters 
in the picture of rough energy landscapes.

\section{Formulation}

Consider the single layer perceptron
with $N\gg 1$ input nodes $\{\xi_j\}$ 
connecting to a single output node by the weights $\{J_j\}$ 
and often, the bias $\theta$ as well. 
For convenience we assume 
that the inputs $\xi_j$ are Gaussian variables with mean 0 and variance 1, 
and the output state is a function $f(x)$ 
of the {\it activation} $x$ at the output node, 
where $x = \vec J\cdot\vec\xi+\theta$. 

The training set consists of $p\equiv\alpha N$ examples 
which map inputs $\{\xi^\mu_j\}$ to the outputs 
$\{S_\mu\}\ (\mu=1,\dots, p)$. 
In the case of random examples, 
$S_\mu$ are random binary variables, 
and the perceptron is used as a storage device. 
In the case of teacher-generated examples, 
$S_\mu$ are the outputs generated by a teacher perceptron 
with weights $\{B_j\}$ and often, a bias $\phi$ as well, 
namely $S_\mu = f(y_\mu)$;
$y_\mu = \vec B\cdot\vec\xi^\mu+\phi$.

Batch learning is achieved by adjusting the weights $\{J_j\}$ iteratively
so that a certain cost function 
in terms of the activations $\{x_\mu\}$ and the output $S_\mu$
of all examples is minimized.
Hence we consider a general cost function $E = -\sum_\mu g(x_\mu,y_\mu)$.
The precise functional form of $g(x,y)$ depends on 
the adopted learning algorithm. 
In previous studies, 
$g(x,y)=-(S-x)^2/2$ in Adaline learning \cite{hertz,opper,krogh}, 
and $g(x,y)=xS$ in Hebbian learning \cite{coolen,rae}. 

To ensure that the perceptron
fulfills the prior expectation of minimal complexity,
it is customary to introduce a weight decay term. 
In the presence of noise, the gradient descent dynamics 
of the weights is given by
\begin{equation}
	{dJ_j(t)\over dt}
	={1\over N}\sum_\mu g'(x_\mu(t),y_\mu)\xi^\mu_j
	-\lambda J_j(t)+\eta_j(t),
\label{original}
\end{equation}
where the prime represents partial differentiation with respect to $x$, 
$\lambda$ is the weight decay strength, 
and $\eta_j(t)$ is the noise term at temperature $T$ with 
\begin{equation}
        \langle\eta_j(t)\rangle=0
        \quad{\rm and}\quad
        \langle\eta_j(t)\eta_k(s)\rangle
        ={2T\over N}\delta_{jk}\delta(t-s).
\label{noise}
\end{equation}
The dynamics of the bias $\theta$ is similar, 
except that no bias decay should be present 
according to consistency arguments \cite{bishop},
\begin{equation}
	{d\theta(t)\over dt}
	={1\over N}\sum_\mu g'(x_\mu(t),y_\mu)+\eta_\theta(t).
\label{bias}
\end{equation}

\section{The Cavity Method}

Our theory is the dynamical version of the cavity method 
\cite{mpv,epl,nips96}. 
It uses a self-consistency argument 
to consider what happens when a new example is added to a training set. 
The central quantity in this method is the {\it cavity activation}, 
which is the activation of a new example 
for a perceptron trained without that example. 
Since the original network has no information about the new example, 
the cavity activation is random.
Here we present the theory for $\theta=\phi=0$,
skipping extensions to biased perceptrons.
Denoting the new example by the label 0,
its cavity activation at time $t$ is
$h_0(t)=\vec J(t)\cdot\vec\xi^0$.
For large $N$, $h_0(t)$ is a Gaussian variable. 
Its covariance is given by the correlation function $C(t,s)$ 
of the weights at times $t$ and $s$, that is,
$\langle h_0(t)h_0(s)\rangle
=\vec J(t)\cdot\vec J(s)\equiv C(t,s)$,
where $\xi^0_j$ and $\xi^0_k$ 
are assumed to be independent for $j\ne k$. 
For teacher-generated examples, the distribution is further specified by 
the teacher-student correlation $R(t)$, given by
$\langle h_0(t)y_0\rangle=\vec J(t)\cdot\vec B\equiv R(t)$.

Now suppose the perceptron incorporates the new example 
at the batch-mode learning step at time $s$. 
Then the activation of this new example at a subsequent time $t>s$ 
will no longer be a random variable. 
Furthermore, the activations of the original $p$ examples at time $t$ 
will also be adjusted from $\{x_\mu(t)\}$ to $\{x^0_\mu(t)\}$ 
because of the newcomer, 
which will in turn affect the evolution of the activation of example 0, 
giving rise to the so-called Onsager reaction effects. 
This makes the dynamics complex, 
but fortunately for large $p\sim N$, 
we can assume that the adjustment 
from $x_\mu(t)$ to $x^0_\mu(t)$ is small, 
and linear response theory can be applied.

Suppose the weights of the original and new perceptron at time $t$ 
are $\{J_j(t)\}$ and $\{J^0_j(t)\}$ respectively. 
Then a perturbation of (\ref{original}) yields
\begin{equation}
	\left({d\over dt}+\lambda\right)(J^0_j(t)-J_j(t))
        ={1\over N}g'(x_0(t),y_0)\xi^0_j
        +{1\over N}\sum_{\mu k} \xi^\mu_j g''(x_\mu(t),y_\mu)
	\xi^\mu_k(J^0_k(t)-J_k(t)).
\label{dyneqn}
\end{equation}
The first term on the right hand side describes the primary effects 
of adding example 0 to the training set, 
and is the driving term for the difference between the two perceptrons. 
The second term describes the many-body reactions 
due to the changes of the original examples caused by the added example,
and is referred to as the Onsager reaction term.
One should note the difference between the cavity and generic activations 
of the added example. 
The former is denoted by $h_0(t)$ 
and corresponds to the activation in the perceptron $\{J_j(t)\}$, 
whereas the latter, denoted by $x_0(t)$ 
and corresponding to the activation in the perceptron $\{J^0_j(t)\}$, 
is the one used in calculating the gradient 
in the driving term of (\ref{dyneqn}).
Since their notations are sufficiently distinct, 
we have omitted the superscript 0 in $x_0(t)$, 
which appears in the background examples $x^0_\mu(t)$.

The equation can be solved by the Green's function technique, yielding 
\begin{equation}
	J^0_j(t)-J_j(t)
	=\sum_k\int ds G_{jk}(t,s)
	\left({1\over N}g'_0(s)\xi^0_k\right),
\label{dressed}
\end{equation}
where $g'_0(s)\equiv g'(x_0(s),y_0)$ 
and $G_{jk}(t,s)$ is the {\it weight Green's function}, 
which describes how the effects of a perturbation 
propagates from weight $J_k$ at learning time $s$ to
weight $J_j$ at a subsequent time $t$.
In the present context, the perturbation comes from 
the gradient term of example 0, 
such that integrating over the history and summing over all nodes 
give the resultant change from $J_j(t)$ to $J^0_j(t)$.

For large $N$ the weight Green's function can be found 
by the diagrammatic approach explained in Appendix A. 
The result is self-averaging 
over the distribution of examples and is diagonal, 
i.e. $\lim_{N\to\infty}G_{jk}(t,s)=G(t,s)\delta_{jk}$, where
\begin{equation}
	G(t,s)=G^{(0)}(t-s)
	+\alpha\int dt_1\int dt_2 G^{(0)}(t-t_1)
	\langle D_\mu(t_1,t_2)g''_\mu(t_2)\rangle G(t_2,s).
\label{wgreen}
\end{equation}
Here the bare Green's function $G^{(0)}(t-s)$ is given by
\begin{equation}
	G^{(0)}(t-s)\equiv\Theta(t-s)\exp(-\lambda(t-s)).
\label{bare}
\end{equation}
$\Theta$ is the step function.
$D_\mu(t,s)$ is the {\it example Green's function} given by
\begin{equation}
	D_\mu(t,s)
	=\delta(t-s)
        +\int dt'D_\mu(t,t')g''_\mu(t')G(t',s).
\label{xgreen}
\end{equation}
Our key to the macroscopic description of the learning dynamics 
is to relate the activation of the examples to their cavity counterparts, 
which is known to be Gaussian. 
Multiplying both sides of (\ref{dressed}) and summing over $j$, we have
\begin{equation}
	x_0(t)-h_0(t)
        =\int ds G(t,s)g'_0(s).
\label{generic}
\end{equation}
In turn, the covariance of the cavity activation distribution 
is provided by the fluctuation-response relation
explained in Appendix B,
\begin{equation}
	C(t,s)
        =\alpha\int dt'G^{(0)}(t-t')\langle g'_\mu(t')x_\mu(s)\rangle
	+2T\int dt'G^{(0)}(t-t')G(s,t').
\label{correlation}
\end{equation}
Furthermore, for teacher-generated examples, 
its mean is related to the teacher-student correlation given by
\begin{equation}
	R(t)
	=\alpha\int dt'G^{(0)}(t-t')\langle g'_\mu(t')y_\mu\rangle.
\label{tscorrelation}
\end{equation}
For a given teacher activation $y$ of a trained example,
the distribution for a set of student activation $\{x(t)\}$
of the same example at different times is,
in the limit of infinitesimal time steps $\Delta t$, given by
\begin{eqnarray}
	&&p(\{x(t)\}|y)
        ={1\over\sqrt{{\rm det}C}}\prod_t\int{dh(t)\over\sqrt{2\pi}}
	\exp\Biggl\{-{1\over 2}\sum_t
    	[h(t)-R(t)y]C(t,s)^{-1}[h(s)-R(s)y]\Biggr\}
	\nonumber\\
	&&\prod_t\delta\left[x(t)-h(t)-\Delta t\sum_sG(t,s)g'(x(s))\right].
\end{eqnarray}
This can be written in an integral form
which is often derived from path integral approaches,
\begin{eqnarray}
	&&p(\{x(t)\}|y)
        =\prod_t\int{dh(t)d\hat h(t)\over 2\pi}
        \exp\Biggl\{i\int dt\hat h(t)[h(t)-R(t)y]
	-{1\over 2}\int dt\int ds
	\hat h(t)C(t,s)\hat h(s)\Biggr\}
	\nonumber\\
	&&\prod_t\delta\left[x(t)-h(t)-\Delta t\sum_sG(t,s)g'(x(s))\right].
\end{eqnarray}
The above distributions and parameters are sufficient
to describe the progress of learning.
Some common performance measures
used for such monitoring purpose include:

(a) {\it Training error} $\epsilon_t$, 
which is the probability of error for the training examples,
and can be determined from the distribution $p(x|y)$
that the student activation of a trained example takes the value $x$
for a given teacher activation $y$ of the same example.

(b) {\it Test error} $\epsilon_{test}$,  
which is the probability of error 
when the inputs $\xi_j^\mu$ of the training examples 
are corrupted by an additive Gaussian noise of variance $\Delta^2$.
This is a relevant performance measure 
when the perceptron is applied to process data 
which are the corrupted versions of the training data. 
When $\Delta^2=0$, the test error reduces to the training error. 
Again, it can be determined from $p(x|y)$, 
since the noise merely adds 
a variance of $\Delta^2C(t,t)$ to the activations.

(c) {\it Generalization error} $\epsilon_g$ for teacher-generated examples,
which is the probability of error for an arbitrary input $\xi_j$ 
when the teacher and student outputs are compared. 
It can be determined from $R(t)$ and $C(t,t)$ since, 
for an example with teacher activation $y$,
the corresponding student activation is a Gaussian
with mean $R(t)y$ and variance $C(t,t)$.

\section{Simulation results}

The success of the cavity approach is illustrated
by the many results presented previously for the Adaline rule
\cite{many,nips99}.
This is a common learning rule and bears resemblance
with the more common back-propagation rule.
Theoretically, its dynamics is particularly convenient for analysis
since $g''(x)=-1$,
rendering the weight Green's function time translation invariant, 
i.e. $G(t,s)=G(t-s)$.
In this case, the dynamics can be solved by Laplace transform.

The closed form of the Laplace solution for Adaline learning
enables us to examine a number of interesting phenomena
in learning dynamics.
For example, an {\it overtraining}
with respect to the generalization error $\epsilon_g$
occurs when the weight decay is not sufficiently strong,
i.e., $\epsilon_g$ attains a minimum at a finite learning time
before reaching a higher steady-state value.
Overtraining of the test error $\epsilon_{test}$ also sets in
at a sufficiently weak weight decay,
which is approximately proportional to the noise variance $\Delta^2$.
We also observe an equivalence
between average dynamics and noiseless dynamics,
namely that a perceptron constructed using the thermally averaged weights
is equivalent to the perceptron obtained at a zero noise temperature.
All these results are well confirmed by simulations.

\begin{figure} [hbt]
\centering
\centerline{\psfig{figure=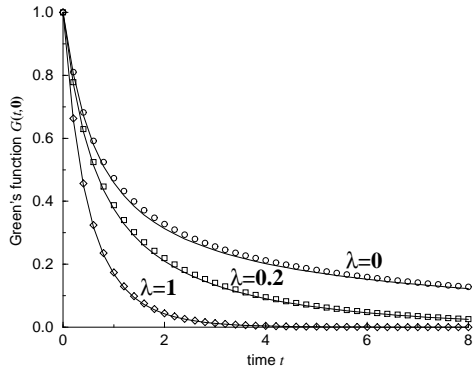,height=5.5cm}}
\vspace{0.5cm}
\caption{
The Green's function $G(t,0)$ for Adaline learning
at a given training set size $\alpha=1.2$ and $T=0$
for different weight decay strengths $\lambda$.
Theory: solid line, simulation: symbols.
}
\label{greenfig}
\end{figure}

Rather than further repeating previous results,
we turn to present results which provide more direct support
to the cavity method.
In the simulational experiment in Fig. \ref{greenfig},
we compare the evolution of two perceptrons
$\{J_j(t)\}$ and $\{J^0_j(t)\}$ in Adaline learning.
At the initial state $J^0_j(0)-J_j(0)=1/N$ for all $j$,
but otherwise their subsequent learning dynamics are exactly identical.
Hence the total sum $\sum_j(J^0_j(t)-J_j(t))$
provides an estimate for the averaged Green's function $G(t,0)$,
which gives an excellent agreement with the Green's function
obtained from the cavity method.

\begin{figure} [hbt]
\centering
\vspace{1cm}
\leftline{\hspace*{1cm}
\psfig{figure=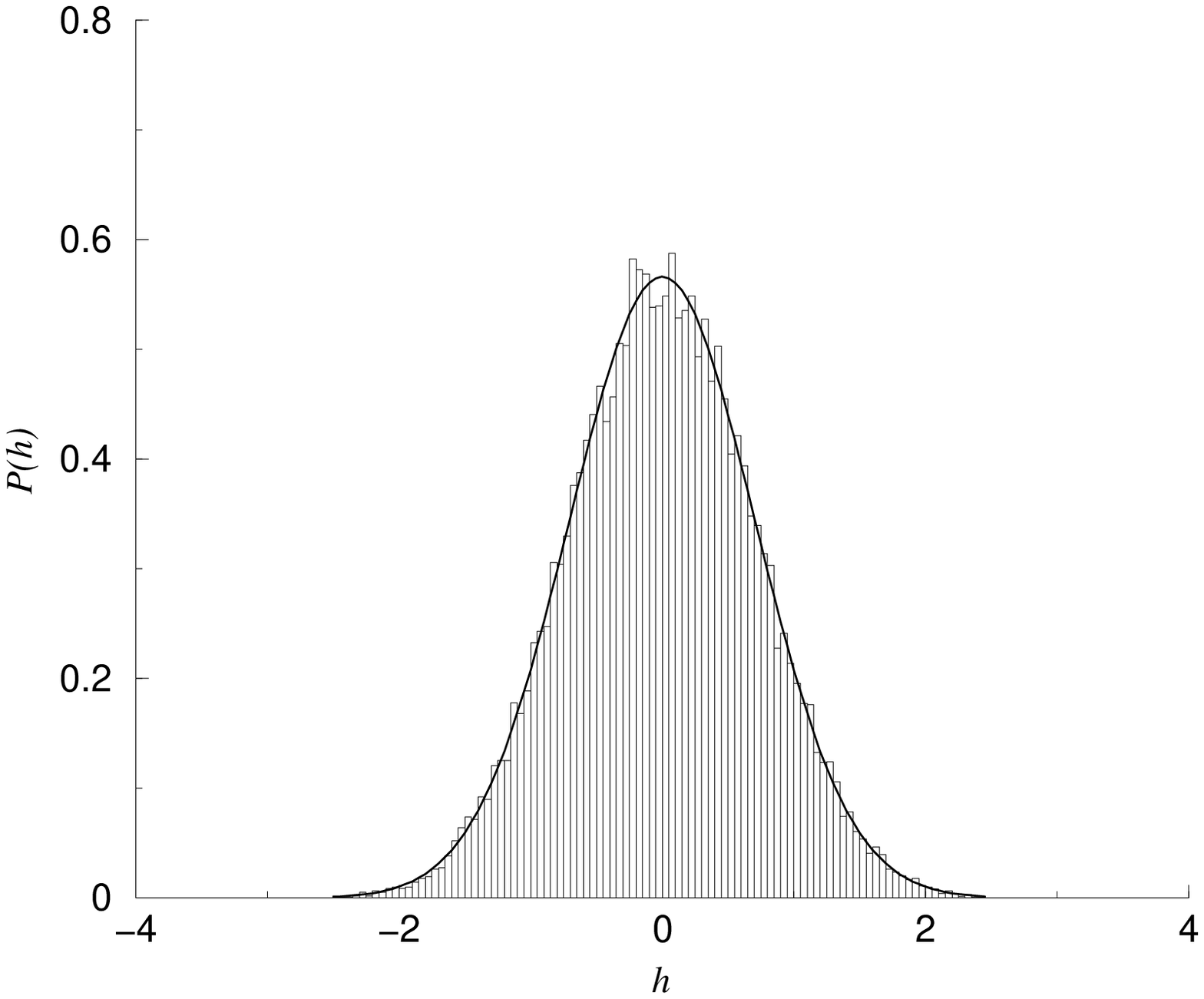,height=5.5cm}}
\vspace{-5.5cm}
\leftline{\hspace*{9cm}
\psfig{figure=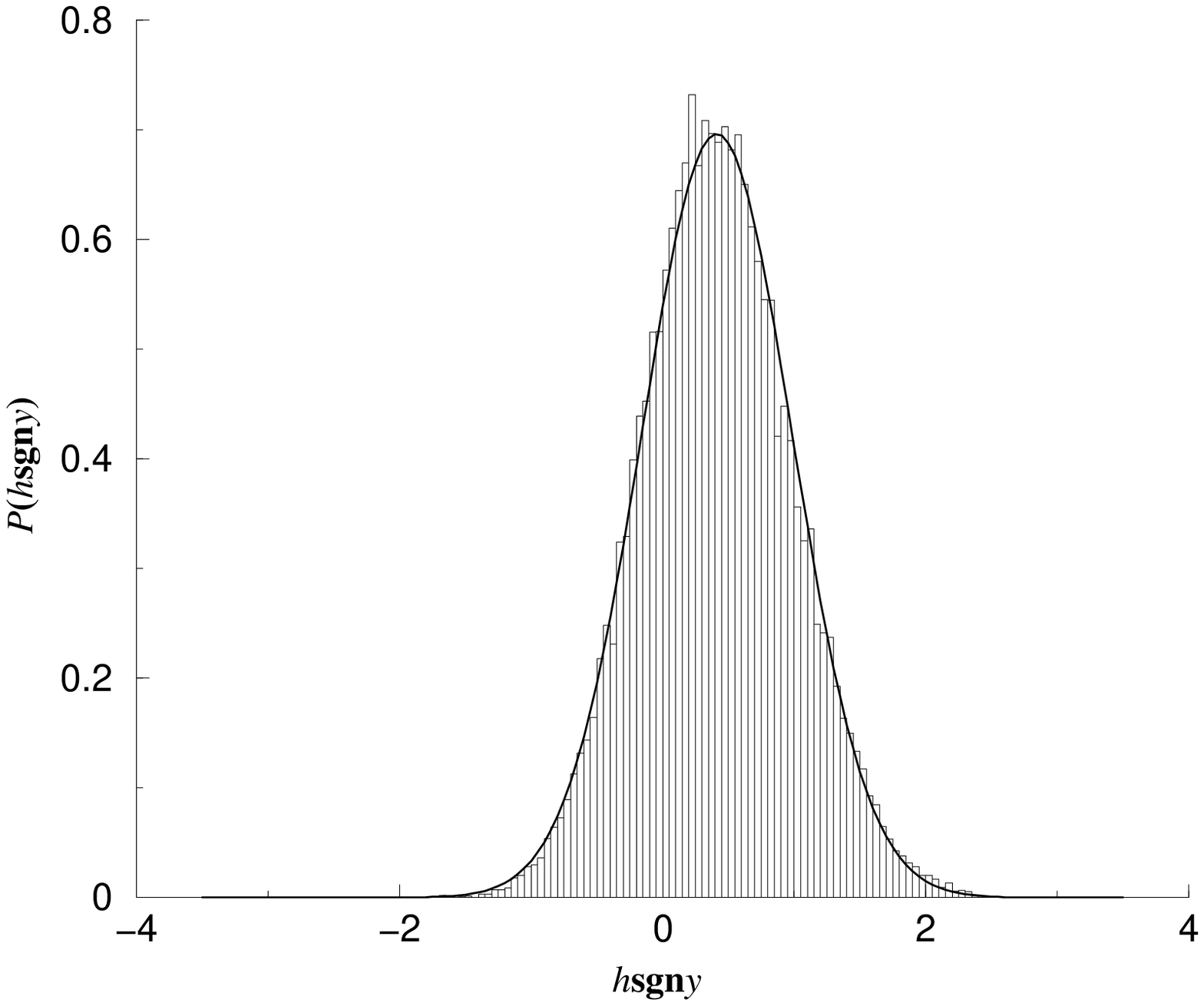,height=5.5cm}}
\vspace{0.5cm}
\caption{
(a) The cavity activation distribution $h$
for Adaline learning at $\alpha=1.2$, $\lambda=0.1$, $T=0$ and $t=2$.
Theory: dashed line, with mean 0 and variance 0.499,
simulation: histogram, with mean 0.000 and variance 0.499.
(b) The distribution of $h{rm sgn}y$. 
Theory: solid line, with mean 0.413 and variance 0.329,
simulation: histogram, with mean 0.416 and variance 0.326.
}
\label{hdisfig}
\end{figure}

Using the Green's function computed from Fig. \ref{greenfig},
we can deduce the cavity activation for each example
by measuring their generic counterpart from the simulation 
and substituting back into Eq. (\ref{generic}).
As shown in the histogram in Fig. \ref{hdisfig}(a),
the cavity activation distribution agrees well
with the Gaussian distribution predicted by the cavity method,
with the predicted mean 0 and variance $C(t,t)$.

Similarly, we show in Fig. \ref{hdisfig}(b)
the distribution of $h{\rm sgn}y$,
i.e., the cavity activation in the direction of the correct teacher output,
The cavity method predicts a Gaussian distribution
with mean $\sqrt{2/\pi}R(t)$ and variance $C(t,t)-2R(t)^2/\pi$.
Again, it agrees well with the histogram obtained from simulation.


\section{Steady-state behavior}

When learning reaches a steady state at $T=0$,
the cavity and generic activations approach a constant.
Hence Eq. (\ref{generic}) reduces to
\begin{equation}
        x_0-h_0=\gamma g'(x_0);
        \quad
        \gamma=\int ds G(t,s),
\label{steady}
\end{equation}
where $\gamma$ is called the local susceptibility in \cite{epl}.
Hence $x_0$ is a well-defined function of $h_0$.

Eq. (\ref{steady}) can also be obtained
by minimizing the change in the steady-state energy function
when example 0 is added,
which is $-g(x_0)+(x_0-h_0)^2/2\gamma$,
the second term being due to
the reaction effects of the background examples.
This was shown in \cite{epl}
for the case of a constant weight magnitude,
but the same could be shown for the case of a constant weight decay.

A self-consistent expression for $\gamma$
can be derived from the steady-state behavior of the Green's function.
Since the system becomes translational invariant in time
at the steady state,
Eqs. (\ref{wgreen}) and (\ref{xgreen}) can be solved by Laplace transform,
yielding
\begin{eqnarray}
	&&\tilde G(z)
        =\tilde G^{(0)}(z)
        +\alpha\tilde G^{(0)}(z)\langle\tilde D_\mu(z)g_\mu''\rangle
        \tilde G(z),
	\\
        &&\tilde D_\mu(z)
        =1+\tilde D_\mu(z)g_\mu''\tilde G(z),
\end{eqnarray}
with $\tilde G^{(0)}(z)=(z+\lambda)^{-1}$.
Identifying $\tilde G(0)$ with $\gamma$, we obtain
\begin{equation}
        \gamma={1\over\lambda}
        +{\alpha\over\lambda}
        \langle{\gamma g_\mu''\over 1-\gamma g_\mu''}\rangle.
\end{equation}
Making use of the functional relation between $x_\mu$ and $h_\mu$,
we have
\begin{equation}
        \gamma={1\over\lambda}(1-\alpha\chi);
        \quad
        \chi=\left\langle 1-{\partial x_\mu\over\partial h_\mu}\right\rangle,
\label{sus}
\end{equation}
where $\chi$ is called the nonlocal susceptibility in \cite{epl}.

At the steady state, the fluctuation response relations
in Eqs. (\ref{correlation}) and (\ref{tscorrelation})
yield the self-consistent equations
for the student-student and teacher-student correlations,
$C\equiv\vec J\cdot\vec J$ and $R\equiv\vec J\cdot\vec B$ respectively,
namely
\begin{equation}
        C={\alpha\over\lambda}\langle g_\mu'x_\mu\rangle;
        \quad
        R={\alpha\over\lambda}\langle g_\mu'y_\mu\rangle.
\end{equation}
Substituting Eqs. (\ref{steady}) and (\ref{sus}),
and introducing the cavity activation distributions, we find 
\begin{eqnarray}
        &&C=(1-\alpha\chi)^{-1}\alpha
        \int Dy\int Dh P(h|y)(x(h)-h)x(h),
        \\
	&&R=(1-\alpha\chi)^{-1}\alpha
        \int Dy\int Dh P(h|y)(x(h)-h)y.
\end{eqnarray}
Since $P(h|y)$ is a Gaussian distribution
with mean $Ry$ and variance $C-R^2$,
its derivatives with respect to $h$ and $R$ are
$-(h-Ry)P(h|y)/(C-R^2)$ and $R(h-Ry)P(h|y)/(C-R^2)$ respectively.
This enables us to use integration by parts
and Eq. (\ref{sus}) for $\chi$ to obtain
\begin{eqnarray}
        &&C=\alpha\int Dy\int Dh P(h|y)(x(h)-h)^2,
        \\
        &&R=\alpha\gamma\int Dy\int Dh P(h|y)
        {\partial x\over\partial h}g_{xy}.
\end{eqnarray}
Hence we have recovered the macroscopic parameters
described by the static version of the cavity method in \cite{epl}
by considering the steady-state behavior of the learning dynamics.
We remark that the saddle point equations in the replica method
also produce identical results,
although the physical interpretation is less transparent
\cite{bouten,bos3}.

We can further derive the microscopic equations
by noting that at equilibrium for $T=0$,
Eq. (\ref{original}) yields
\begin{equation}
        J_j={1\over\lambda N}\sum_\mu g'_\mu\xi^\mu_j,
\end{equation}
which leads to the set of equations
\begin{equation}
        x_\nu={1\over\lambda}\sum_\mu g'_\mu Q_{\mu\nu};
        \quad
        Q_{\mu\nu}\equiv{1\over N}\sum_j\xi^\mu_j\xi^\nu_j.
\label{micro}
\end{equation}
The TAP equations are obtained
by expressing these equations in terms of the cavity activations 
via Eq. (\ref{steady}),
\begin{equation}
        h_\nu
        =\sum_{\mu\ne\nu}(x(h_\mu)-h_\mu)Q_{\mu\nu}
        +\alpha\chi x(h_\nu).
\label{tap}
\end{equation}
The iterative solution of the equation set
was applied to the maximally stable perceptron,
which yielded excellent agreement with the cavity method,
provided that the stability condition
discussed below is satisfied \cite{epl}.
However, the agreement is poorer
when applied to the committee tree \cite{nips96}
and the pruned perceptron \cite{tanc},
where the stability condition is not satisfied.

To study the stability condition of the cavity solution,
we consider the change in the steady-state solution
when example 0 is added to the training set.
Consider the magnitude of the displaced weight vector
$\Delta\equiv\sum_j(J^0_j-J_j)^2$.
Using either the static or dynamic version of the cavity method,
we can show that
\begin{equation}
        \Delta={1\over N}{(x_0-h_0)^2\over
        1-\alpha\left\langle\left(1-{\partial x_\mu\over\partial h_\mu}
        \right)^2\right\rangle}.
\end{equation}
In order that the change due to the added example is controllable,
the stability condition is thus
\begin{equation}
        \alpha\left\langle\left(1-{\partial x_\mu\over\partial h_\mu}
        \right)^2\right\rangle<1.
\end{equation}
This is identical to the stability condition
of the replica-symmetric ansatz in the replica method,
the so-called Almeida-Thouless condition \cite{ws}.

As a corollary, when a band gap exists in the activation distribution,
the stability condition is violated.
This is because the function $x(h)$ becomes discontinuous in this case,
implying the presence of a delta-function component
in $\partial x/\partial h$.

\begin{figure} [hbt]
\centering
\leftline{\hspace*{1cm}
\psfig{figure=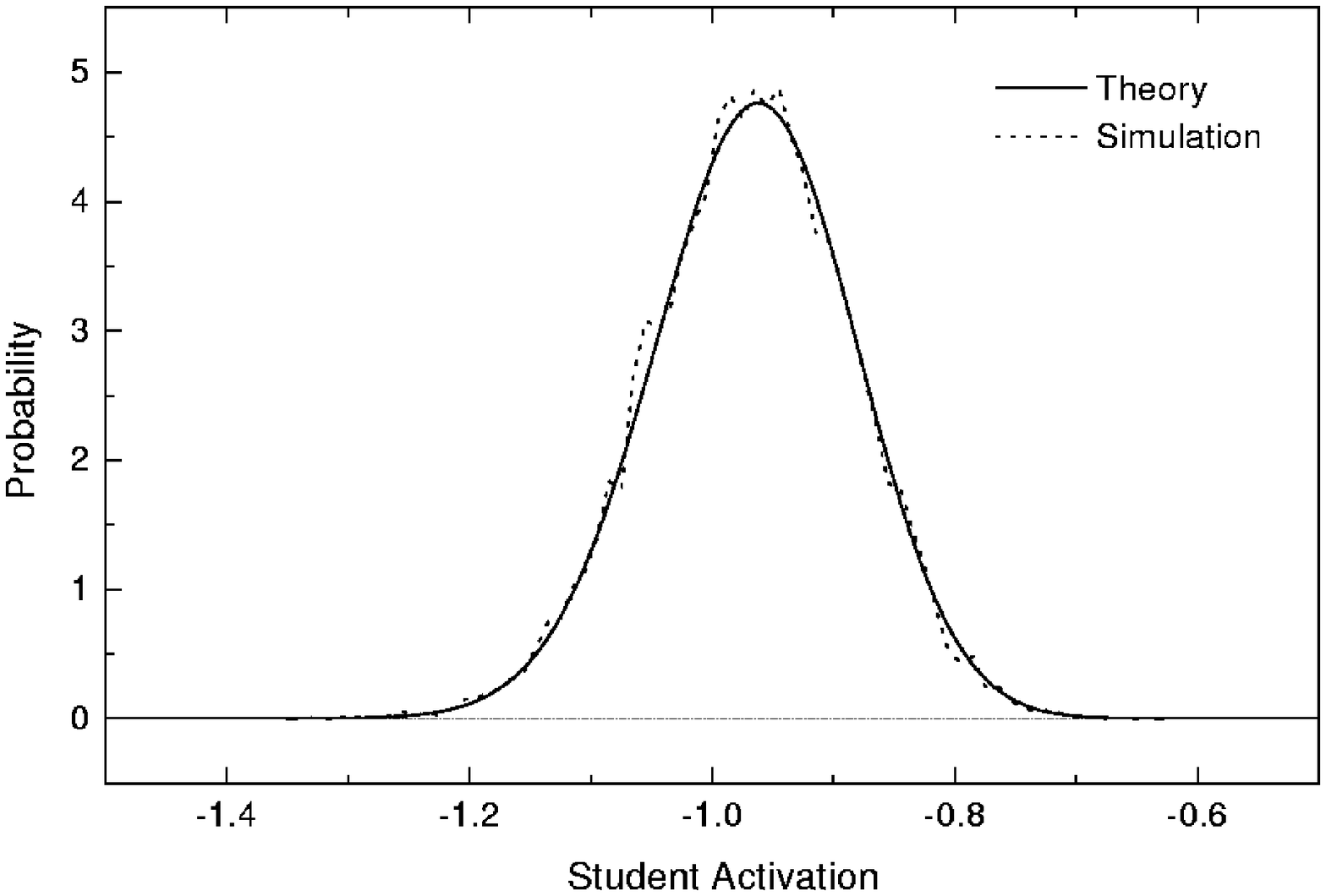,height=5.5cm}}
\vspace{-5.5cm}
\leftline{\hspace*{9cm}
\psfig{figure=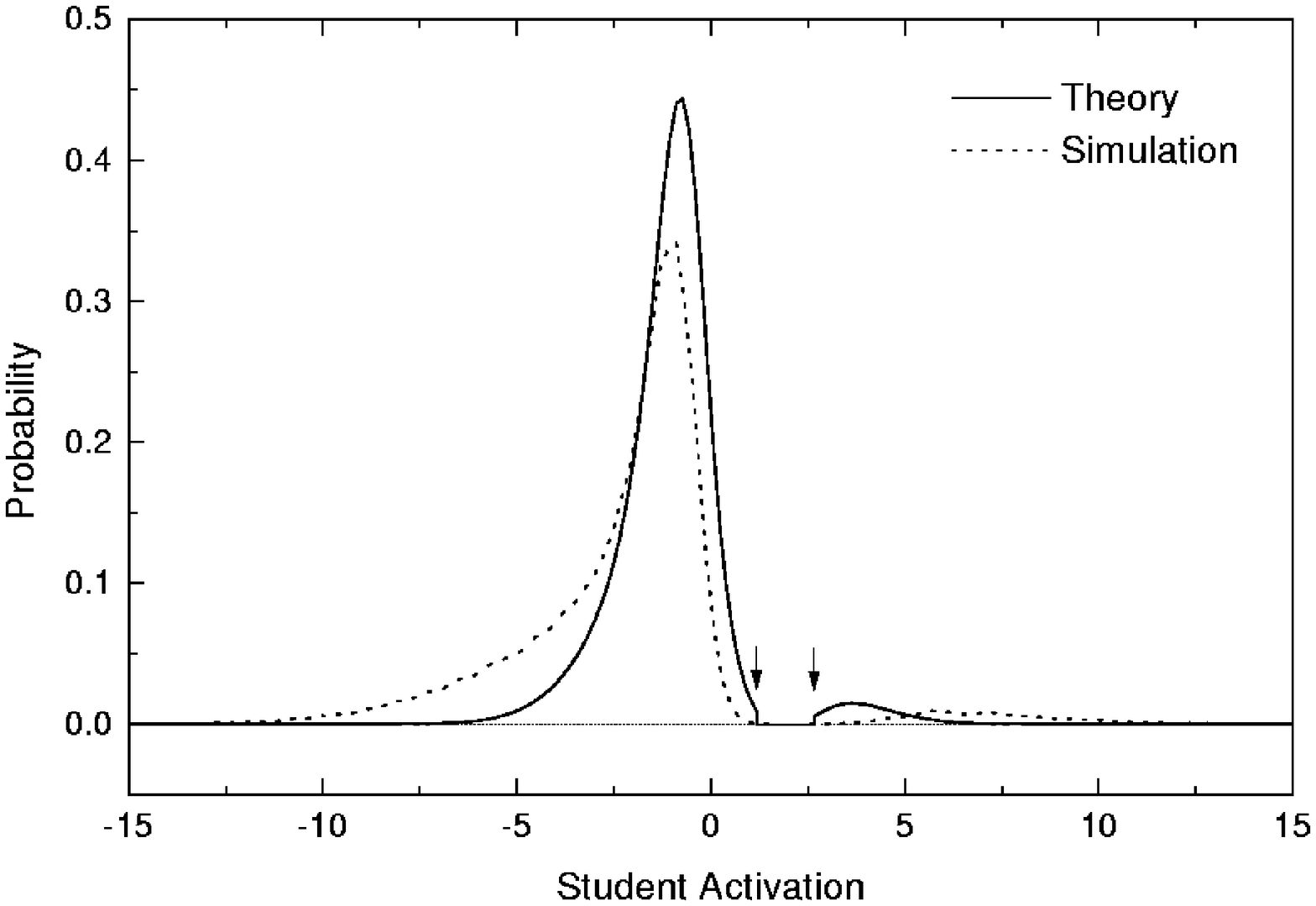,height=5.5cm}}
\vspace{0.5cm}
\caption{
Typical student activation distributions at $\alpha=3$ and $\lambda=0.002$, 
(a) in the stable regime in which 
the teacher activations are corrupted by noises of variance 0.1, 
(b) in the unstable regime in which 
the teacher activations are corrupted by noises of variance 5. 
}
\label{adisfig}
\end{figure}

Such is the case in the nonlinear perceptron trained with noisy examples 
using the backpropagation algorithm \cite{luo}.
For insufficient examples and weak weight decay,
the activation distribution exhibits a gap
for the more difficult examples,
i.e., when the teacher output $y$ and the cavity activation $h$
has a large difference.
As shown in Fig. \ref{adisfig}(a), 
simulational and theoretical predictions of the activation distributions
agree well in the stable regime,
but the agreement is poor in the unstable regime
shown in Fig. \ref{adisfig}(b).
Hence the existence of band gaps necessitates
the picture of a rough energy landscape,
as described in the following section.

\section{The picture of rough energy landscapes}

\begin{figure} [hbt]
\centering
\centerline{\psfig{figure=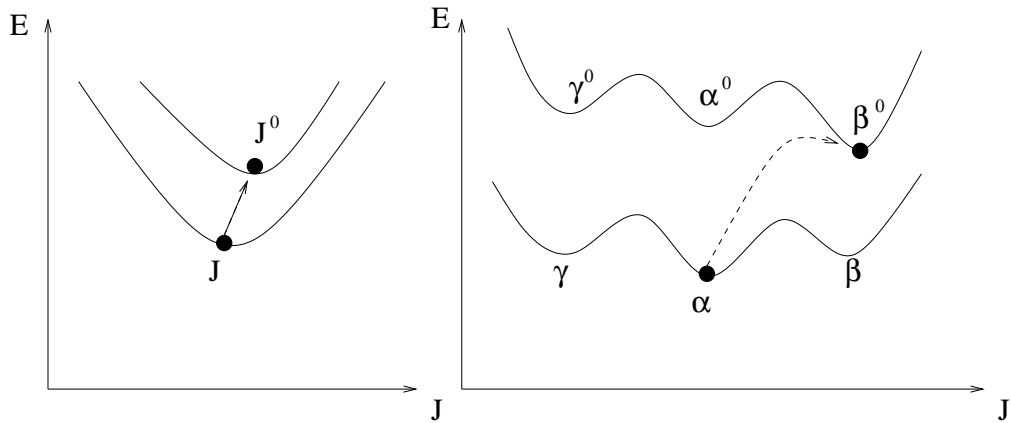,height=5.5cm}}
\vspace{0.5cm}
\caption{
Schematic drawing of the change in the energy landscape
in the weight space when example 0 is added,
for the regime of (a) smooth energy landscape,
(b) rough energy landscape.
}
\label{roughfig}
\end{figure}

To consider what happens beyond the stability regime, 
one has to take into account
the rough energy landscape of the learning space.
To keep the explanation simple,
we consider the learning of examples generated randomly, 
the case of teacher-generated examples
being similar though more complicated.
Suppose that the original global minimum
for a given training set is $\alpha$.
In the picture of a smooth energy landscape,
the network state shifts perturbatively after adding example 0,
as schematically shown in Fig. \ref{roughfig}(a).
In contrast, in the picture of a rough energy landscape,  
a nonvanishing change to the system is induced,
and the global minimum shifts to the neighborhood 
of the local minimum $\beta$,
as schematically shown in Fig. \ref{roughfig}(b).
Hence the resultant activation $x_0^\beta$ 
is no longer a well-defined function
of the cavity activation $h_0^\alpha$.
Instead it is a well-defined function
of the cavity activation $h_0^\beta$.
Nevertheless, one may expect
that correlations exist between the states
$\alpha$ and $\beta$.

Let $q_0$ be the correlation between two local minima
labelled by $\beta$ and $\gamma$,
i.e. $\vec J^\beta\cdot\vec J^\gamma=q_0$. 
Both of them are centred about the global minimum $\alpha$, so that
$\vec J^\alpha\cdot\vec J^\beta
=\vec J^\alpha\cdot\vec J^\gamma=\sqrt{q_0q_1}$,
where $q_1
=\vec J^\alpha\cdot\vec J^\alpha
=\vec J^\beta \cdot\vec J^\beta
=\vec J^\gamma\cdot\vec J^\gamma$.
Since both states $\alpha$ and $\beta$ are determined
in the absence of the added example 0, the correlation
$\langle h_0^\alpha h_0^\beta\rangle=\sqrt{q_0q_1}$ as well.
Knowing that both $h_0^\alpha$ and $h_0^\beta$
obey Gaussian distributions,
the cavity activation distribution can be determined 
if we know the prior distribution of the local minima.

At this point we introduce the central assumption in the cavity method 
for rough energy landscapes: 
we assume that the number of local minima at energy $E$ 
obeys an exponential distribution
\begin{equation}
        d\aleph(E)\propto\exp(-wE)dE.
\end{equation}
Similar assumptions have been used 
in specifying the density of states in disordered systems \cite{mpv}.
Thus the cavity activation distribution is given by
\begin{equation}
        P(h_0^\beta|h_0^\alpha)=
        {G(h_0^\beta|h_0^\alpha)\exp[-w\Delta E(x(h_0^\beta))]\over
        \int dh_0^\beta 
        G(h_0^\beta|h_0^\alpha)\exp[-w\Delta E(x(h_0^\beta))]},
\label{hbeta}
\end{equation}
where $G(h_0^\beta|h_0^\alpha)$ is a Gaussian distribution
with mean $\sqrt{q_0/q_1}h_0^\alpha$ and variance $q_1-q_0$.
$\Delta E$ is the change in energy due to the addition of example 0,
and is equal to $-g(x_0^\beta)+(x_0^\beta-h_0^\beta)^2/2\gamma$.
The weights $J_j^\beta$ are given by
\begin{equation}
        J_j^\beta={1\over\lambda N}
        \sum_\mu g'(x_\mu^\beta)\xi_j^\mu.
\end{equation}
Self-consistent equations for the macroscopic parameters
are derived in Appendix C.
The results are identical to 
the first step replica symmetry-breaking solution
in the replica method.

It remains to check whether the microscopic equations have been modified 
due to the roughening of the energy landscape.
In terms of the generic activations,
the microscopic equations are identical to Eq. (\ref{micro})
for each local minimum.
In terms of the cavity activations, 
the TAP equations are again identical to Eq. (\ref{tap}),
except that the nonlocal susceptibility $\chi$
is now evaluated in the corresponding local minimum.
The cavity activation distribution is no longer a Gaussian distribution,
but is modified by the density of states in Eq. (\ref{hbeta}) now.
Hence the values of $\chi$ and $\gamma$ appearing in the TAP equations
are no longer identical to the case
of restricting learning to a single valley.

\section{Conclusion}

In summary, we have introduced a general framework
for modeling the dynamics of learning based on the cavity method, 
which is applicable to general learning cost functions,
though its tractable solutions are not generally available.

We have verified its validity
by simulations of the cavity activation distributions.
The steady-state behavior is seen to be consistent
with the static version of the cavity method
in the picture of smooth energy landscapes,
which is equivalent to the replica symmetric ansatz in the replica method.
This picture is based on the assumption
that the dynamics is stable against perturbations,
and is manifested in a stability condition
equivalent to the Almeida-Thouless condition in the replica method.
Beyond the stability regime,
rough energy landscapes have to be introduced,
but the microscopic TAP equations remain valid.

There are two interesting issues
concerning the extension of the present work.
First, it is interesting to consider
how the dynamics is modified in the picture of rough energy landscapes.
In this case, aging effects may appear,
and the dynamics may not be translationally invariant
in time \cite{aging}.
Second, it is interesting to consider 
whether the analysis remains tractable for nonlinear learning rules. 
In general, $D_\mu(t,s)$ in (\ref{xgreen}) has to be expanded as a series. 
Nevertheless, we have shown that the asymptotic dynamics
remains tractable for nonlinear learning rules.
For transient dynamics, we may need to consider appropriate approximations.
Another applicable area 
is the case of batch learning with very large learning steps, 
whose analysis remains simple due to its fast convergence \cite{bos}.
The method can also be applied 
to on-line learning of restricted sets of examples.

An alternative general theory for learning dynamics 
is the dynamical replica theory \cite{coolen}. 
It yields exact results for Hebbian learning, 
but for less trivial cases, 
the analysis is approximate and complicated by the need 
to solve replica saddle point equations at every learning instant.
It is hoped that by adhering to an exact formalism, 
the cavity method can provide more fundamental insights 
when extended to multilayer networks.

We thank A. C. C. Coolen and D. Saad for fruitful discussions. 
This work was supported by the Research Grant Council of Hong Kong 
(HKUST6130/97P and HKUST6157/99P).

\appendix
\section{The Green's function}

\begin{figure}[hbt]
\centering
\centerline{\psfig{figure=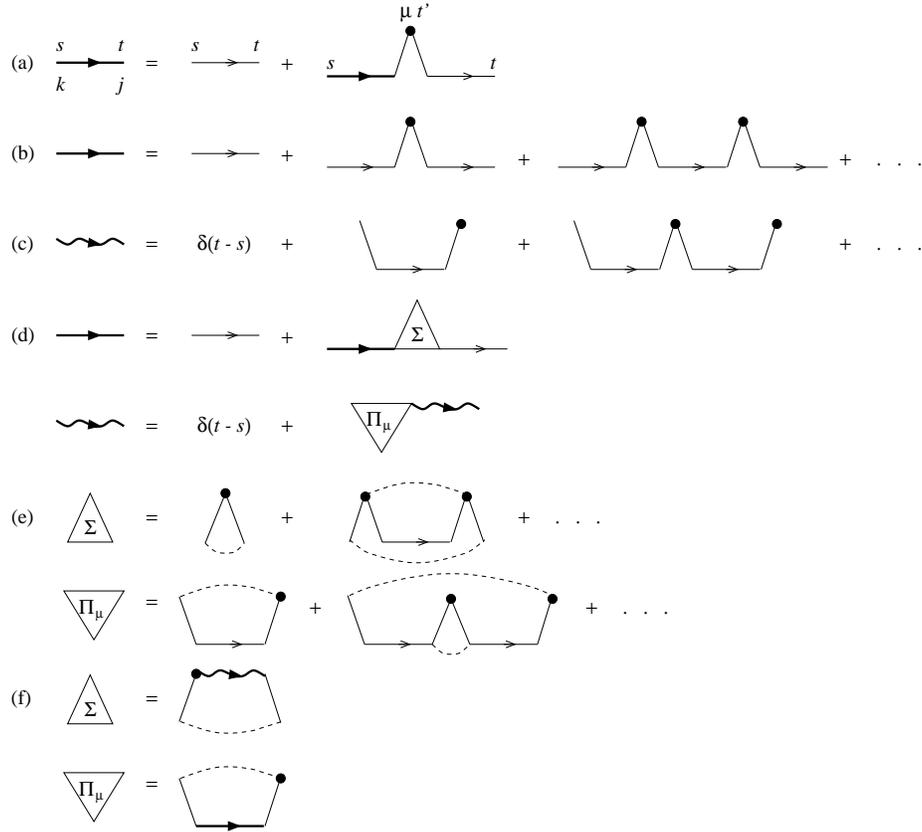,height=11cm}}
\vspace{0.5cm}
\caption{
(a) Diagrammatic representation of Eq. (\ref{diagram});
(b) iterative solution to Eq. (\ref{diagram});
(c) the example Green's function;
(d) Dyson's equations;
(e) the self-energies;
(f) the self-energies in terms of the Green's functions.
}
\label{diagfig}
\end{figure}

Substituting Eq. (\ref{dressed}) into Eq. (\ref{dyneqn}),
we see that the Green's function satisfies
\begin{equation}
        \left({d\over dt}+\lambda\right)G_{jk}(t,s)
        =\delta(t-s)\delta_{jk}
        +{1\over N}\sum_{\mu i} \xi^\mu_j g''_\mu(t)
        \xi^\mu_iG_{ik}(t,s).
\end{equation}
Introducing the bare Green's function $G^{(0)}(t-s)$ in Eq. (\ref{bare}),
\begin{equation}
        G_{jk}(t,s)=G^{(0)}(t-s)\delta_{jk}
        +{1\over N}\sum_{\mu i} \int dt' G^{(0)}(t-t')
        \xi^\mu_j g''_\mu(t')
        \xi^\mu_iG_{ik}(t',s).
\label{diagram}
\end{equation}
This equation is represented diagrammatically in Fig. \ref{diagfig}(a).
We use a slanted line to represent an example bit,
the top and bottom ends of the line
corresponding to the example label and node label respectively.
A filled circle represents $g''_\mu(t)$.
Thin and thick lines represent the bare and dressed Green's functions
$G^{(0)}(t-s)$ and $G(t,s)$ respectively.
The iterative solution to Eq. (\ref{diagram})
can be represented by the series of diagrams in Fig. \ref{diagfig}(b).
It is convenient to concurrently introduce
the {\it example} Green's function $D_\mu(t,s)$
as shown in Fig. \ref{diagfig}(c).

The average over the distribution of example inputs
is done by pairing of example or node labels
and are represented by dashed lines connecting the vertices 
above or below the solid lines.
Pairing of example and node labels
yield factors of 1 and $\alpha$ respectively.
Noting that crossing diagrams do not contribute \cite{hertz},
the two Green's functions can be expressed
in terms of the self-energies $\Sigma$ and $\Pi_\mu$,
via the Dyson's equations in Fig. \ref{diagfig}(d).
The self-energies are defined in Fig. \ref{diagfig}(e),
and are characterized by having the first node or example
paired with the last one only.
The self-energies can in turn be expressed
in terms of the Green's functions as in Fig. \ref{diagfig}(f),
thus allowing for self-consistent solutions.

After eliminating the self-energies,
the results of the diagrammatic analysis are given by
Eqs. (\ref{wgreen}) and (\ref{xgreen}).

\section{The fluctuation response relation}

In terms of the bare Green's function,
the solution to the dynamical equation Eq. (\ref{original}) is
\begin{equation}
        J_j(t)
        ={1\over N}\sum_\mu \int dt' G^{(0)}(t-t') g'_\mu(t') \xi^\mu_j
        +\int dt' G^{(0)}(t-t') \eta_j(t').
\label{weight}
\end{equation}
Multiplying both sides by $J_j(s)$ and summing over $j$, we have
\begin{equation}
        C(t,s)
        =\alpha \int dt' G^{(0)}(t-t')
        \langle g'_\mu(t') x_\mu(s)\rangle
	+\int dt' G^{(0)}(t-t') \sum_j J_j(s) \eta_j(t').
\end{equation}
The correlation between $J_j(s)$ and $\eta_j(t')$
can be considered by comparing the learning process with another one
which is noiseless between $t'-\epsilon$ and $t'+\epsilon$,
but is otherwise identical.
Denoting the weight of this alternative process 
by $J_j^{\backslash\eta(t')}$, we have
\begin{equation}
        J_j(s)
        =J_j^{\backslash\eta(t')}(s)
        +\int_{t'-\epsilon}^{t'+\epsilon} dt'' G(s,t'') \eta_j(t'').
\end{equation}
Noting that $J_j^{\backslash\eta(t')}(s) $ is uncorrelated with $\eta_j(t')$,
and $\eta_j(t'')$ has a delta function correlation with $\eta_j(t')$ 
as in Eq. (\ref{noise}), 
we arrive at Eq. (\ref{correlation}).

Similarly, multiplying both sides by $B_j$ and summing over $j$,
we arrive at Eq. (\ref{tscorrelation}).

\section{Macroscopic parameters in rough energy landscapes}

From Eq. (\ref{sus}), the nonlocal susceptibility is given by
\begin{equation}
        \chi=\int dh_0^\alpha G(h_0^\alpha)
        {\int dh_0^\beta G(h_0^\beta|h_0^\alpha)e^{-w\Delta E}
        (1-\partial x_0^\beta/\partial h_0^\beta)\over
        \int dh_0^\beta G(h_0^\beta|h_0^\alpha)e^{-w\Delta E}},
\label{chi}
\end{equation}
where $G(h_0^\alpha)$ is a Gaussian with mean 0 and variance $q_1$.
The local susceptibility $\gamma$ is given by
\begin{equation}
        \gamma={1\over\lambda(1-\alpha\chi)}.
\label{gamma}
\end{equation}
From the fluctuation response relation in Eq. (\ref{correlation}), we have
\begin{equation}
        q_1={\alpha\over\lambda}\int dh_0^\alpha G(h_0^\alpha)
        {\int dh_0^\beta G(h_0^\beta|h_0^\alpha)e^{-w\Delta E}
        g'(x_0^\beta)x_0^\beta\over
        \int dh_0^\beta G(h_0^\beta|h_0^\alpha)e^{-w\Delta E}},
\end{equation}
Substituting Eqs. (\ref{steady}) and (\ref{sus}), we find
\begin{equation}
        (1-\alpha\chi)q_1=\alpha\int dh_0^\alpha G(h_0^\alpha)
        {\int dh_0^\beta G(h_0^\beta|h_0^\alpha)e^{-w\Delta E}
        (x_0^\beta-h_0^\beta)x_0^\beta)\over
        \int dh_0^\beta G(h_0^\beta|h_0^\alpha)e^{-w\Delta E}}.
\end{equation}
The differentiations of $G(h_0^\beta|h_0^\alpha)$
with respect to $h_0^\beta$ and $h_0^\alpha$ introduce factors of 
$-(h_0^\beta-\sqrt{q_0/q_1}h_0^\alpha)/(q_1-q_0)$ and
$\sqrt{q_0/q_1}(h_0^\beta-\sqrt{q_0/q_1}h_0^\alpha)/(q_1-q_0)$ respectively,
and that of $G(h_0^\alpha)$ with respect to $h_0^\alpha$
introduces $-h_0^\alpha/q_1$.
This allows us to use integration by parts
and Eq. (\ref{sus}) for $\chi$ to obtain
\begin{eqnarray}
        &&q_1=
	\alpha\left[1+{w\over\gamma}(q_1-q_0)\right]
	\int dh_0^\alpha G(h_0^\alpha)
        {\int dh_0^\beta G(h_0^\beta|h_0^\alpha)e^{-w\Delta E}
        (x_0^\beta-h_0^\beta)^2\over
        \int dh_0^\beta G(h_0^\beta|h_0^\alpha)e^{-w\Delta E}}
        \nonumber\\
        &&+\alpha{w\over\gamma}q_0
        \int dh_0^\alpha G(h_0^\alpha)\Biggl\{
        {\int dh_0^\beta G(h_0^\beta|h_0^\alpha)e^{-w\Delta E}
        (x_0^\beta-h_0^\beta)^2\over
        \int dh_0^\beta G(h_0^\beta|h_0^\alpha)e^{-w\Delta E}}
	-\left[
        {\int dh_0^\beta G(h_0^\beta|h_0^\alpha)e^{-w\Delta E}
        (x_0^\beta-h_0^\beta)\over
        \int dh_0^\beta G(h_0^\beta|h_0^\alpha)e^{-w\Delta E}}
        \right]^2\Biggr\}.
\label{q1}
\end{eqnarray}
Next we derive an equation for the interstate overlap $q_0$.
Consider the steady-state solution of a local minimum $J_j^\beta$
given by Eq. (\ref{steady}).
Multiplying both sides by the weight vector $J_j^\gamma$
at another local minimum and summing over $j$, we have
\begin{equation}
        q_0={1\over\lambda N}\sum_\mu g'(x_\mu^\beta)x_\mu^\gamma.
\end{equation}
Proceeding as in the case of $q_1$, we get
\begin{eqnarray}
        &&q_0=\alpha\left[1+{w\over\gamma}(q_1-q_0)\right]
	\int dh_0^\alpha G(h_0^\alpha)\Biggl[
        {\int dh_0^\beta G(h_0^\beta|h_0^\alpha)e^{-w\Delta E}
        (x_0^\beta-h_0^\beta)\over
        \int dh_0^\beta G(h_0^\beta|h_0^\alpha)e^{-w\Delta E}}
	\Biggr]^2
	\nonumber\\
	&&+\alpha{w\over\gamma}q_0
        \int dh_0^\alpha G(h_0^\alpha)\Biggl\{
        {\int dh_0^\beta G(h_0^\beta|h_0^\alpha)e^{-w\Delta E}
        (x_0^\beta-h_0^\beta)^2\over
        \int dh_0^\beta G(h_0^\beta|h_0^\alpha)e^{-w\Delta E}}
	-\left[
        {\int dh_0^\beta G(h_0^\beta|h_0^\alpha)e^{-w\Delta E}
        (x_0^\beta-h_0^\beta)\over
        \int dh_0^\beta G(h_0^\beta|h_0^\alpha)e^{-w\Delta E}}
        \right]^2\Biggr\}.
\label{q0}
\end{eqnarray}
Solving Eqs. (\ref{q1}) and (\ref{q0}),
\begin{eqnarray}
        &&\int dh_0^\alpha G(h_0^\alpha)
        {\int dh_0^\beta G(h_0^\beta|h_0^\alpha)e^{-w\Delta E}
        (x_0^\beta-h_0^\beta)^2\over
        \int dh_0^\beta G(h_0^\beta|h_0^\alpha)e^{-w\Delta E}}
	={q_1+{w\over\gamma}(q_1-q_0)^2\over
        \alpha\left[1+{w\over\gamma}(q_1-q_0)\right]^2},
\label{qq1}\\
        &&\int dh_0^\alpha G(h_0^\alpha)\left[
        {\int dh_0^\beta G(h_0^\beta|h_0^\alpha)e^{-w\Delta E}
        (x_0^\beta-h_0^\beta)\over
        \int dh_0^\beta G(h_0^\beta|h_0^\alpha)e^{-w\Delta E}}
        \right]^2
	={q_0\over\alpha\left[1+{w\over\gamma}(q_1-q_0)\right]^2}.
\label{qq0}
\end{eqnarray}
To determine the distribution of local minima, 
namely the parameter $w$, 
we introduce a ``free energy'' $F(p, N)$
for $p$ examples and $N$ input nodes, given by
\begin{equation}
        d\aleph(E)=\exp[w(F(p, N)-E)]dE.
\label{dos}
\end{equation}
This ``free energy'' determines the averaged energy of the local minima 
and should be an extensive quantity, i.e. it should scale as the system size. 
Cavity arguments enable us to find an expression $F(p+1, N)-F(p, N)$. 
When the number of examples increases by 1,
the density of states for a given $h_0^\alpha$ are related by
\begin{equation}
        \aleph(E_{p+1},h_0^\alpha)
        =\int dE_p\aleph(E_p,h_0^\alpha)
        \int dh_0^\beta G(h_0^\beta|h_0^\alpha)
        \delta(E_{p+1}-E_p-\Delta E).
\end{equation}
Using Eq. (\ref{dos}) we obtain, on averaging over $h_0^\alpha$,
\begin{equation}
        F(p+1,N)
        =F(p,N)-{1\over w}\int dh_0^\alpha G(h_0^\alpha)
        \ln\int dh_0^\beta G(h_0^\beta|h_0^\alpha)e^{-w\Delta E}.
\end{equation}
Similarly, we may consider a cavity argument
for the addition of one input node, 
expanding the network size from $N$ to $N+1$. 
Skipping the details, the final result is
\begin{equation}
        F(p,N+1)-F(p,N)
	=-{q_0\over 2\gamma\left[1+{w\over\gamma}(q_1-q_0)\right]}
        -{1\over 2w}\ln\left[1+{w\over\gamma}(q_1-q_0)\right]
        +{\lambda\over 2}q_1.
\end{equation}
Since $F$ is an extensive quantity, $F(p, N)$ should scale as $N$ 
for a given ratio $\alpha=p/N$. This implies
\begin{equation}
	{F\over N}
        ={\partial F\over\partial N}
        =(F(p, N+1)-F(p, N))+\alpha(F(p+1, N)-F(p, N)).
\end{equation}
When $E=F$, the density of states reduces to $O(e^0)$
and the global minimum is reached. Hence
\begin{eqnarray}
        &&\int dh_0^\alpha G(h_0^\alpha)
        {\int dh_0^\beta G(h_0^\beta|h_0^\alpha)e^{-w\Delta E}
        g(x_0^\beta)\over
        \int dh_0^\beta G(h_0^\beta|h_0^\alpha)e^{-w\Delta E}}
	={q_0\over 2\gamma\left[1+{w\over\gamma}(q_1-q_0)\right]}
	\nonumber\\
	&&+{1\over 2w}\ln\left[1+{w\over\gamma}(q_1-q_0)\right]
	+{\alpha\over w}\int dh_0^\alpha G(h_0^\alpha)
        \ln\int dh_0^\beta G(h_0^\beta|h_0^\alpha)e^{-w\Delta E}.
\label{entropy}
\end{eqnarray}
Eqs. (\ref{chi}), (\ref{gamma}), (\ref{qq1}), (\ref{qq0}) and (\ref{entropy})
form a set of five equations for $\chi$, $\gamma$, $q_1$, $q_0$ and $w$.

\end{document}